# Radio Astronomy transformed

Aperture Arrays – Past, Present and Future

Michael A. Garrett
Netherlands Institute for Radio Astronomy (ASTRON)
Postbus 2, 7990 AA Dwingeloo
The Netherlands
garrett@astron.nl

*Abstract*— Aperture Arrays have played a major role in radio astronomy since the field emerged from the results of long-distance communication tests performed by Karl Jansky in the early 1930's. The roots of this technology extend back beyond Marconi, although the first electronically scanned instrument only appeared in the run-up to World War II. After the war, phased arrays had a major impact in many walks of life, including astronomy and astrophysics. Major progress was made in understanding the nature of the radio sky, including the discovery of Pulsars. Despite these early successes, parabolic dishes largely replaced aperture arrays through the 1960's, and right up until the end of the 20$^{th}$ century. Technological advances in areas such as signal processing, digital electronics, low-power/high performance super-computing and large capacity data storage systems have recently led to a substantial revival in the use of aperture arrays – especially at frequencies below 300 MHz. Composed of simple antennas with commercially available low-noise room-temperature amplifiers, aperture arrays with huge collecting areas can be synthesized at relatively low cost. Multiple beams (or multiple fields-of-view) can be rapidly formed and electronically steered across the sky. As astronomers begin to grapple with these new possibilities, the next goal is to see these systems move to higher GHz frequencies. Aperture Arrays operating at frequencies of up to 1.7 GHz are expected to form a substantial part of the Square Kilometre Array (SKA).

## I. Introduction

For more than 100 years, advances in Radio Science have impacted almost every aspect of human activity – in work, rest and play. Radio astronomy itself is a pursuit that originally arose from experiments designed to improve long distance radio communications [1]. Large phased arrays such as LOFAR [2] currently dominate low-frequency radio astronomy (< 300 MHz) and a programme is now underway to extend the range of these systems to GHz frequencies through the international Square Kilometre Array project [3]. In this paper, I lightly review the early adoption of large antenna arrays in radio astronomy, their subsequent demise and their resurrection a generation later.

## II. The First Antenna Arrays

Large aperture arrays, as currently employed by radio astronomers at low frequencies have their roots in the large antenna arrays originally deployed by Marconi in his first attempts at transatlantic wireless communication.

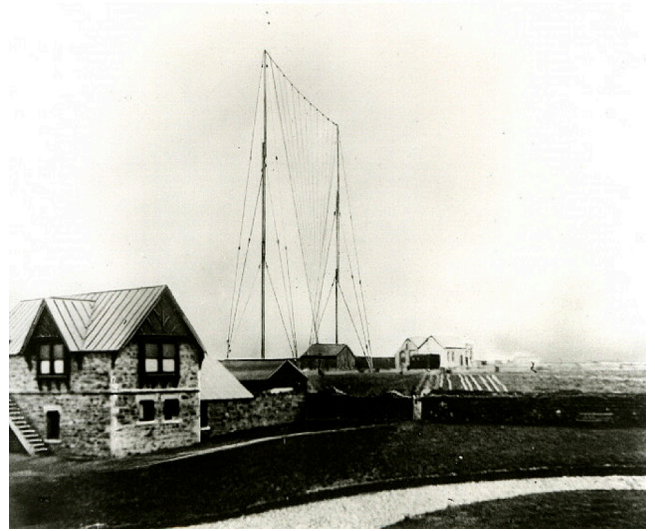

Fig.1. Marconi's antenna array at Poldhu, Cornwall. From here the first radio signals were successfully transmitted across the Atlantic to North America.

Marconi understood that not only was transmission power important in these early experiments but so too was antenna directivity and gain [4]. As early as 1902, a British engineer, Sidney Brown had connected multiple antennas together in order to achieve directional gain in both transmission and reception – a concept that he patented [5]. Soon after inventing the ground antenna, Marconi built large centrally fed antenna arrays that were comparable in scale with the antenna stations now used for radio astronomy (see Fig.1). While the established scientific world looked on skeptically at his early work, the role played by "Marconi wireless rooms'' in saving hundreds of lives during the sinking of the RMS Titanic soon changed their minds. Marconi had been "lucky" in some senses - it was only later that the role of the ionosphere in reflecting low frequency radio waves beyond the line-of-sight horizon was understood to be crucial in these early successes.

The first phased arrays can be traced back to the early work of the German physicist Karl Ferdinand Braun [6]. Braun who incidentally shared the Nobel Prize in Physics with Marconi in 1909, constructed the first switchable phased array (see Fig.2) with a three element system that formed an equilateral triangle.

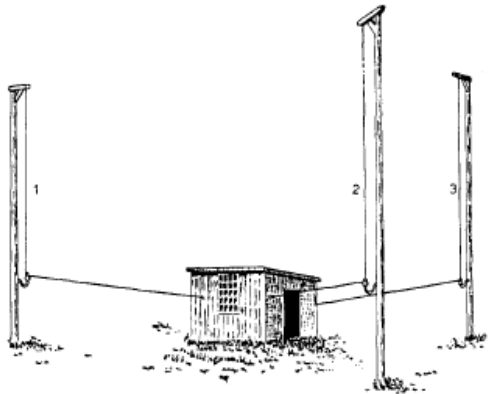

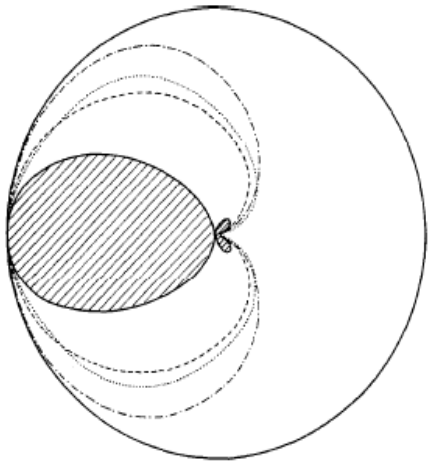

Fig.2. a 3 element antenna system constructed by Braun (1909) in which an extra delay could be inserted into one of the elements in order to change the gain of the centrally fed array.

A cable delay ($\lambda/4$) could be switched into any element's path, permitting the gain of the array to be rapidly rotated around fixed intervals of 120 degrees. It wasn't until just before the outbreak of the second world war that continuous phased arrays systems (see Fig.3) were demonstrated by Friis[1] & Feldmanin [7] Despite these exciting developments, radar systems used during the WWII were much more conservative and were typically fixed or only pointed by mechanically steering the array. The 'Chain Home' (CH) radar systems deployed by Britain are good examples of fixed systems. Capable of detecting enemy aircraft at a distance of several hundred kilometres [8], these became operational just in time for the Battle of Britain (1940), and their success almost certainly changed the course of the war.

III. PHASED ARRAYS FOR RADIO ASTRONOMY

After the war, many radar scientists and engineers in Australia, Great Britain & the USA returned to civilian life. On its own, Britain's Telecommunications Research Establishment (TRE) furnished several future leaders in the field of radio astronomy such as Bolton, Bowen, Hanbury Brown, Hey, Hewish, Lovell & Ryle. Electronically steerable phased arrays began to find application in many different areas. For example, at the Rad Lab at MIT, Luis Alvarez built an electronically steered dipole array that became the basis for both the first Approach and Landing System (ALS) and post-WWII early warning systems.

In radio astronomy, phased array systems were to make some of the most important discoveries in the field. These included the detection of radio emission from the Sun and Jupiter; the first lunar and planetary radar measurements, the first interferometer experiments, the detection of meteorite trails with velocities bounded by the solar system escape velocity, plus the first systematic surveys of the sky revealing new phenomena such as radio galaxies, quasars and pulsars, to name only a few (see [9] for the early history of radio astronomy). Large synthesised telescopes like the Cambridge interferometer (also known as the "4-acre array") also built by Hewish, discovered pulsars and produced the 2C and 3C catalogues. These were the first radio instruments to employ the new technique of aperture synthesis and also boasted up to 16 independent beams, well separated in declination. In addition, the first radio telescope with an aperture encompassing 1 square kilometre was also realised by Reber via a huge array of dipoles operating at 2 MHz. Though never fully exploited, Reber's array was certainly ahead of its time in realising the potential of cheaply deploying large-scale apertures at low radio frequencies.

IV. THE RISE OF PARABOLIC DISHES AND DEMISE OF APERTURE ARRAYS

Despite the major role played by aperture arrays in the early days of radio astronomy, the prominence of these systems sharply decreased from the early 1960's onwards. From this point, almost all major investments in telescope instrumentation were typically realised in the form of parabolic dishes. An early post-war example includes the 25-metre Dwingeloo Telescope – the first large, fully steerable antenna for radio astronomy. Until the end of the 20th century, parabolic dishes were set to dominate radio astronomy at cm, mm, and sub-mm wavelengths.

---
[1] Interestingly, Friis was also one of the team involved in developing the receiver system for Jansky's original discovery of cosmic radio emission at Bell Labs in New Jersey.

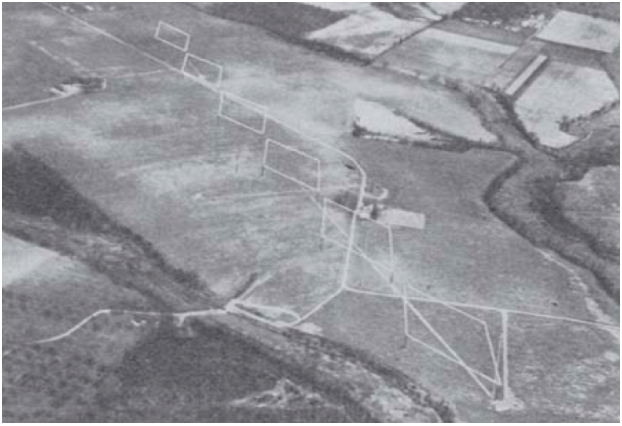

Fig.3. Aerial view of Friis and Feldman's Multiple Unit Steerable Antenna (1937).

It is interesting to consider the reasons for this sea change in the field. Certainly, steel parabolas offered the ability to move radio astronomy up to much higher frequencies. This permitted higher resolution measurements to be made which greatly aided the astronomer's task of cross-matching radio sources with higher resolution optical catalogues. The fact that the extended emission from the Milky Way also decreased at higher frequencies, aided the study of the discrete sources in the extra-galactic sky. Since the prediction of the hyperfine spectral line associated with neutral hydrogen by van der Hulst in 1944, there was also a clear scientific drive to map out the Milky Way and other Galaxies by observing the 21cm (1.4 GHz) line. Dishes also offered greater reliability – the substantial and often complex cabling of phased arrays often led to problems that were often difficult to locate quickly. With a single focal point, highly specialised low-noise receivers could also be used with dishes – this greatly increased the sensitivity of the system via cryogenically cooling of the related amplifier systems.

## V. REINVENTING RADIO ASTRONOMY – APERTURE ARRAYS RE-DISCOVERED

In the early 1990s, staff at ASTRON, the Netherlands Institute for Radio Astronomy, began thinking about the feasibility of building a huge radio telescope with a collecting area of about 1 square kilometre ([10]) – Aperture Arrays began to seem interesting again. Arnold van Ardenne, then Director of ASTRON's R&D labs had recently returned from Ericsson in 1994, and set up an aperture array development line [11], mostly focused on systems operating at much higher frequencies than before –1-2 GHz (see Fig. 4). This led to the production of a series of aperture array tiles: the AAD (Adaptive Antenna Demonstrator), OSMA (One Square Metre Array), THEA (THousand Element Array) and subsequently EMBRACE (Electronic Multi-beam Radio Astronomy ConcEpt) as part of the international FP6 SKADS project [12]. In the slipstream of these developments, the concept of LOFAR also began to gain traction [13,14,15].

Technological advances in areas such as signal processing, digital electronics, low-power/high performance super-computing and large capacity data storage systems meant that this new generation of aperture array telescopes would be quite unlike anything that had gone before. The concept of the "Software Telescope" arose, as it was realised that today's electronics permitted many copies of signals to be made at the individual dipole or tile level - independent multiple beams (multiple fields-of-view) could be easily created and rapidly shifted across the sky. High-speed networks, based on new optical fibre technologies also permitted distributed real-time arrays to be constructed, separated by hundreds or even thousands of kilometres. By buffering data at the dipole level via the availability of large, inexpensive RAM modules, retrospective imaging of the entire sky also became possible. In short, the ICT revolution had transformed aperture arrays from an unwieldy, unreliable and forgotten technology of yesteryear, to the basis for a new kind of telescope with the potential to completely transform radio astronomy.

## VI. RADIO ASTRONOMY TRANSFORMED

Modern Aperture Array telescopes [16], offer many advantages over conventional parabolic systems. Aperture arrays focus incoming radiation by varying the delay (or more often phase) electronically across the aperture. Dishes focus light via reflection from the parabolic metal surface. An aperture array provides a fully unblocked aperture with an unrestricted view of the entire sky. Composed of simple antennas with commercially available low-noise room-temperature amplifiers, huge collecting areas can be synthesized at relatively low cost. Multiple beams (or multiple fields-of-view) can be rapidly formed and electronically steered across the sky. Retrospective imaging of the sky is possible by buffering raw voltage data at the individual dipole level, and later recombining these data from the entire array with the appropriate delay (i.e. the desired pointing direction). In short, a modern aperture array system provides the ultimate flexibility, reliability and performance with no moving parts involved.

Since the turn of the 21$^{st}$ century, aperture arrays have reappeared as major elements of the next generation of low-frequency radio telescopes. Fuelled by technological advances and the growing interest to study neutral hydrogen in the early universe, telescopes such as LOFAR [2], the MWA, PAPER and the LWA [17] are currently under construction or being commissioned. LOFAR is by far the largest example with the broadest range of observing modes. With baseline lengths extending from the North of the Netherlands, through Germany, Sweden, France & the UK, sub-arcsecond resolution is possible at the highest LOFAR observing frequencies (~ 200 MHz). Fig. 5 shows LOFAR and the other low-frequency aperture array telescopes that have emerged over the last few years. The first commissioning results from these telescopes are impressive, clearly demonstrating the huge scientific potential of aperture array systems.

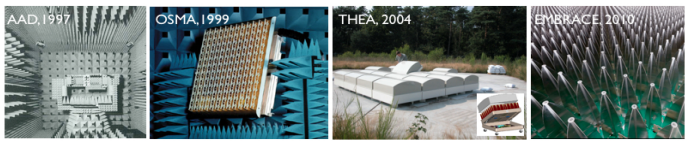

Fig.4. The evolution of ASTRON's aperture array programme over the last 15 years.

The revival in aperture array technology development, also led to the realisation that the field of view of traditional parabolic dishes could also be transformed via the introduction of Phased Array Feeds (PAFs). These are essentially aperture arrays placed at the focus of a dish. Typically a PAF will extend the field of view of a traditional parabolic dish by a factor of about 30. The ability to form telescope beams using the weighted sum of all the PAF antenna elements also permits very high efficiencies to be achieved via optimal illumination of the dish. Since it is impossible to cool hundreds of amplifiers, this improved efficiency largely compensates for the higher system temperatures of PAFs due to the use of room temperature amplifiers. Nevertheless, significant improvements in room-temperature LNAs continue to be made and this trend is expected to continue for the next few years.

These PAFs can be thought of as a 2-dimensional receiver array – something like a "radio camera" with 100s of "pixels" populating the focal plane [18]. In 2006, modified versions of ASTRON's THEA tile (see Fig. 4) were successfully tested first at the focus of CSIRO's Fleurs dish in Sydney and then at one of the WSRT antennas. Based on these early successes, large PAF-based telescope systems like ASKAP and the WSRT-APERTIF are now under construction.

Astronomers are conservative – at least some of them are – and while many have been slow to re-embrace this "old" technology (with a few notable exceptions – see Section 2.1), the rest of the world has adopted aperture array systems in many different guises. The same advantages that aperture arrays offer to radio astronomy (large and multiple field-of-views, rapid electronic steering, reliability, flexibility, cost and performance) are not surprisingly, also considered important in other fields. Aperture arrays also offer some fields other distinct advantages - in the realm of military operations their flat profiles are much easier to seamlessly incorporate into the surface structure of other systems (e.g. vehicles, ships, aircraft fuselage and missile skins) than parabolic dishes. Notably, these systems are typically operating at 1-10 GHz. Military systems are also known to include significant redundancy and "self-healing" features (referred to as "smart skins") in which the systems can be automatically reconfigured after being partially disabled. Aperture Arrays now replace dish systems for both active and passive radar applications (civilian and military systems), weather monitoring, surveillance, navigation and communications (both ground and space). Even in areas that are often reticent to take up new technologies, aperture arrays are present e.g. NASA's Messenger spacecraft, now in orbit around Mercury is the first deep space mission to employ aperture arrays as its main communication system. In the future, the further adoption of so-called steerable "smart antennas" (aperture arrays) in mass-market wireless applications is expected to have a huge impact on the world of mobile and wireless communications [19].

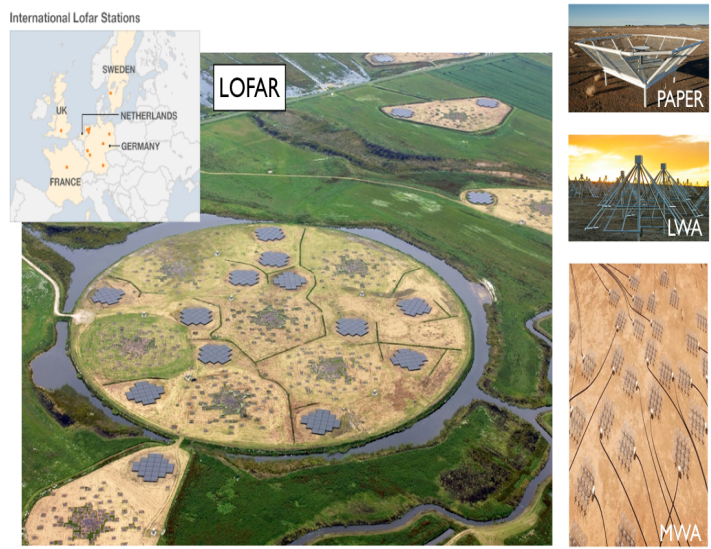

Fig.5. Radio astronomy reinvented - the 21st century has seen an explosion of aperture array telescopes being constructed – LOFAR (left), the MWA, LWA & PAPER.

All these new developments will surely have an important impact on the form and design of the SKA, especially on the longer timescales of SKA Phase 2.

I. THE FUTURE OF APERTURE ARRAYS AND THE SKA

Aperture arrays operating at frequencies between 0.1 and 1.5 GHz are naturally one of the major enabling elements that can make the SKA a truly transformational telescope, well into the middle of this century. The study of neutral hydrogen in the local, distant and early universe was the original driver for the SKA, and it still remains the premier scientific case. Phase 1 includes a low-frequency array (SKA1-LO) that naturally builds on pathfinders such as LOFAR and precursors such as the MWA in Western Australia. Phase 2 is envisaged to include a major Dense Aperture Array (DAA) component, operating at frequencies of ~ 0.5-1.5 GHz to be sited in South Africa.

There is a clear need to continue to advance DAA tile developments and to construct a significant precursor telescope before embarking on the construction of a much larger DAA in Phase 2. Currently known as "EMMA" [20], we propose the construction of a 16-station SKA-2 precursor array becoming operational before the end of the decade (see Fig. 6). This ambitious timescale should obviously be considered in the context of the latest SKA timeline. The SKA must be an instrument that will amaze future generations to come! Building a SKA that is simply the "VLA on steroids" is simply not good enough – we have the ability to do much, *much* better. The present generation of astronomers, and the generations to come, deserve nothing less.

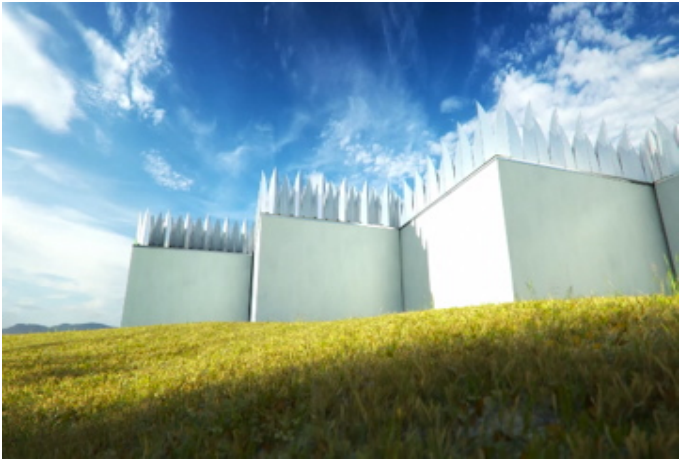

Fig.6. Precursors utilising dense aperture array technology are a necessary step before they are introduced as a major element of SKA (Phase 2).

ACKNOWLEDGMENT *(Heading 5)*

I'd like to acknowledge the staff of ASTRON and our international partners for re-introducing Aperture Arrays into the field of radio astronomy – chiefly via the LOFAR, SKADS, MWA, LWA & PAPER projects.